# AN ANALYTICAL FRAMEWORK FOR DATA STREAM MINING TECHNIQUES BASED ON CHALLENGES AND REQUIREMENTS

MAHNOOSH KHOLGHI

Department of Electronic, Computer and IT,
Islamic Azad University, Qazvin Branch, Qazvin, Iran
and member of Young Researchers Club
m.kholghi@qiau.ac.ir

MOHAMMADREZA KEYVANPOUR

Department of Computer Engineering
Alzahra University Tehran, Iran
keyvanpour@alzahra.ac.ir

Abstract:
A growing number of applications that generate massive streams of data need intelligent data processing and online analysis. Real-time surveillance systems, telecommunication systems, sensor networks and other dynamic environments are such examples. The imminent need for turning such data into useful information and knowledge augments the development of systems, algorithms and frameworks that address streaming challenges. The storage, querying and mining of such data sets are highly computationally challenging tasks. Mining data streams is concerned with extracting knowledge structures represented in models and patterns in non stopping streams of information. Generally, two main challenges are designing fast mining methods for data streams and need to promptly detect changing concepts and data distribution because of highly dynamic nature of data streams. The goal of this article is to analyze and classify the application of diverse data mining techniques in different challenges of data stream mining. In this paper, we present the theoretical foundations of data stream analysis and propose an analytical framework for data stream mining techniques.

*Keywords: Data Stream, Data Stream Mining, Stream Preprocessing.*

1. Introduction

Data mining techniques are suitable for simple and structured data sets like relational databases, transactional databases and data warehouses. Fast and continuous development of advanced database systems, data collection technologies, and the World Wide Web, makes data grow rapidly in various and complex forms such as semi-structured and non-structured data, spatial and temporal data, and hypertext and multimedia data. Therefore, mining of such complex data becomes an important task in data mining realm. In recent years different approaches are proposed to overcome the challenges of storing and processing of fast and continuous streams of data [5, 27].

Data stream can be conceived as a continuous and changing sequence of data that continuously arrive at a system to store or process [7]. Imagine a satellite-mounted remote sensor that is constantly generating data. The data are massive (e.g., terabytes in volume), temporally ordered, fast changing, and potentially infinite. These features cause challenging problems in data streams field. Traditional OLAP and data mining methods typically require multiple scans of the data and are therefore infeasible for stream data applications [20]. Whereby data streams can be produced in many fields, it is crucial to modify mining techniques to fit data streams. Data stream mining has many applications and is a hot research area. With recent progress in hardware and software technologies, different measurement can be done in various fields. These measurements are continuously feasible for data with high changing ratio. Common applications which require mining of large amount of data to find new patterns are sensor networks, store and search of web events, and computer networks traffic. These patterns are valuable for decision makings.

Data Stream mining refers to informational structure extraction as models and patterns from continuous data streams. Data Streams have different challenges in many aspects, such as computational, storage, querying and mining. Based on last researches, because of data stream requirements, it is necessary to design new techniques to replace the old ones. Traditional methods would require the data to be first stored and then processed off-line using complex algorithms that make several pass over the data, but data stream is infinite and data generates with high rates, so it is impossible to store it. Therefore two main challenges are [4, 10]:

- designing fast mining methods for data streams and;
- need to detect promptly changing concepts and data distribution because of highly dynamic nature of data streams

A first research challenge is designing fast and light mining methods for data streams, for example, algorithms that only require one pass over the data and work with limited memory. Another challenge is created by the highly dynamic nature of data streams, whereby the stream mining algorithms need to detect promptly changing concepts and data distribution and adapt to them.

Based on the type of data stream mining challenges in this paper we propose a comprehensive classification of data stream mining challenges and then evaluate new methods of data stream mining. In this evaluation we consider the relation of these methods with different data mining techniques in an analytical manner. The rest of this paper is structured as follow. Section 2 reviews data stream mining algorithms and presents a classification of these algorithms based on their dependencies to pre-processing techniques. In section 3, a classification of data stream mining challenges and current proposed approaches is presented. Section 4 presents and discusses our analytical framework and section 5 presents our conclusions and directions for future works.

## 2. Data stream mining

High volume and potential infinite data streams are generated bye So many resources such as real-time surveillance systems, communication networks, Internet traffic, on-line transactions in the financial market or retail industry, electric power grids, industry production processes, scientific and engineering experiments, remote sensors, and other dynamic environments. In data stream model, data items can be relational tuples like network measurements and call records. In comparison with traditional data sets, data stream flows continuously in systems with varying update rate. Data streams are continuous, temporally ordered, fast changing, massive and potentially infinite. Due to huge amount and high storage cost, it is impossible to store an entire data streams or to scan through it multiple times. So it makes so many challenges in storage, computational and communication capabilities of computational systems. Because of high volume and speed of input data, it is needed to use semi-automatic interactional techniques to extract embedded knowledge from data.

Data stream mining is the extraction of structures of knowledge that are represented in the case of models and patterns of infinite streams of information. The general process of data stream mining is depicted in Fig. 1.

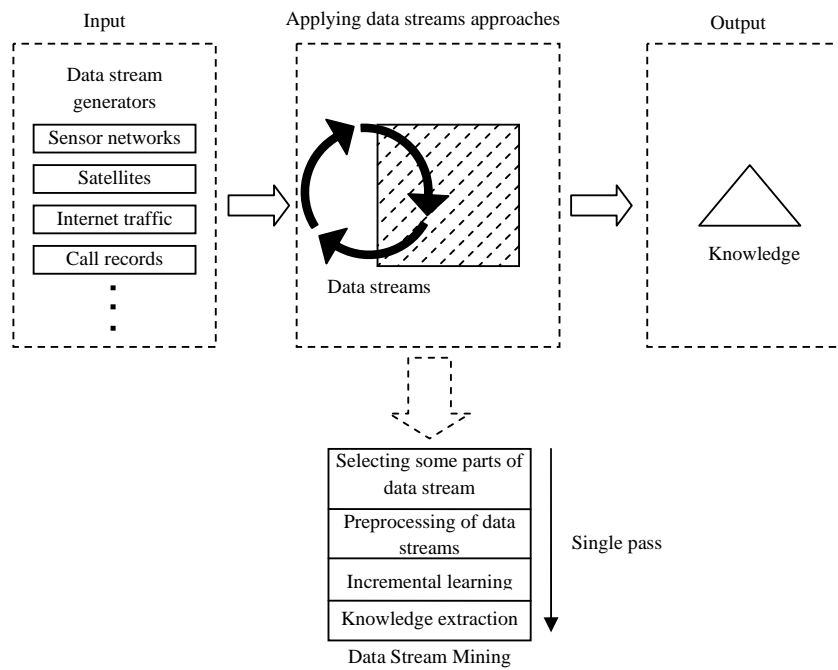

Fig. 1. General process of data stream mining

For extracting knowledge or patterns from data streams, it is crucial to develop methods that analyze and process streams of data in multidimensional, multi-level, single pass and online manner. These methods should not be limited to data streams only, because they are also needed when we have large volume of data. Moreover, because of the limitation of data streams, the proposed methods are based on statistic, calculation and complexity theories. For example, by using summarization techniques that are derived from statistic science, we can confront with memory limitation. In addition, some of the techniques in computation theory can be used for implementing time and space efficient algorithms. By using these techniques we can also use common data mining approaches by enforcing some changes in data streams [27].

Some solutions have been proposed based on data stream mining problems and challenges. These solutions can be categorized to data-based and task-based solutions. This classification is depicted in Fig. 2. Data-based techniques refer to summarizing the whole dataset or choosing a subset of the incoming stream to be analyzed. Sampling, load and sketching techniques represent the former one. Synopsis data structures and aggregation represent the later one. Task-based techniques are those methods that modify existing techniques or invent new ones in order to address the computational challenges of data stream processing. Approximation algorithms, sliding window and algorithm output granularity represent this category.

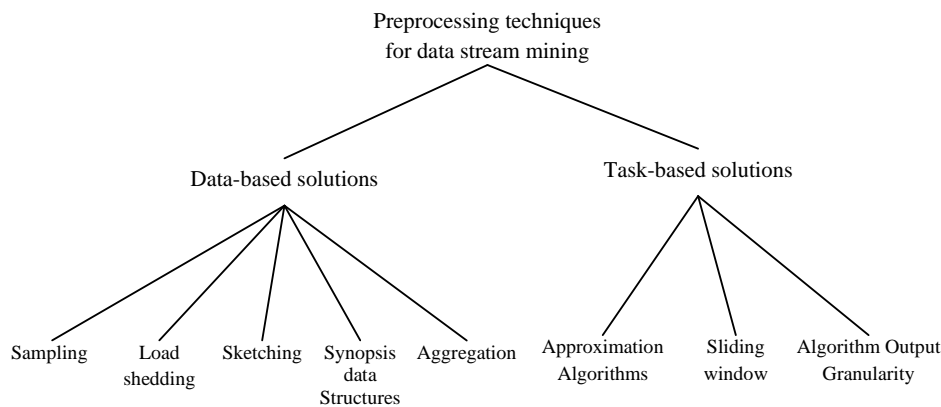

Fig. 2. Classification of data stream preprocessing methods

Sampling refers to the process of probabilistic choice of a data item to be processed or not. The problem with using sampling in the context of data stream analysis is the unknown dataset size. Thus the treatment of data stream should follow a special analysis to find the error bounds. Another problem with sampling is that it would be important to check for anomalies for surveillance analysis as an application in mining data streams. Sampling may not be the right choice for such an application. Sampling also does not address the problem of fluctuating data rates. It would be worth investigating the relationship among the three parameters: data rate, sampling rate and error bounds.

Load shedding refers to the process of dropping a sequence of data streams. Load shedding has been used successfully in querying data streams. It has the same problems of sampling. Load shedding is difficult to be used with mining algorithms because it drops chunks of data streams that could be used in the structuring of the generated models or it might represent a pattern of interest in time series analysis [6, 30].

Sketching is the process of randomly project a subset of the features. It is the process of vertically sample the incoming stream. Sketching has been applied in comparing different data streams and in aggregate queries. The major drawback of sketching is that of accuracy. It is hard to use it in the context of data stream mining [5, 27].

Creating synopsis of data refers to the process of applying summarization techniques that are capable of summarizing the incoming stream for further analysis. Wavelet analysis, histograms, quantiles and frequency moments have been proposed as synopsis data structures. Since synopsis of data does not represent all the characteristics of the dataset, approximate answers are produced when using such data structures.

The process in which the input stream is represented in a summarized form is called aggregation. This aggregate data can be used in data mining algorithms. The main problem of this method is that highly fluctuating data distributions reduce the method's efficiency [1-3].

Approximation algorithms have their roots in algorithm design. It is concerned with design algorithms for computationally hard problems. These algorithms can result in an approximate solution with error bounds. The idea is that mining algorithms are considered hard computational problems given its features of continuality and speed and the generating environment that is featured by being resource constrained.

Approximation algorithms have attracted researchers as a direct solution to data stream mining problems. However, the problem of data rates with regard with the available resources could not be solved using approximation algorithms. Other tools should be used along with these algorithms in order to adapt to the available resources. Approximation algorithms have been used in [27].

The inspiration behind sliding window is that the user is more concerned with the analysis of most recent data streams. Thus the detailed analysis is done over the most recent data items and summarized versions of the old ones.

The algorithm output granularity (AOG) introduces the first resource-aware data analysis approach that can cope with fluctuating very high data rates according to the available memory and the processing speed represented in time constraints. The AOG performs the local data analysis on a resource constrained device that generates or receive streams of information. AOG has three main stages. Mining followed by adaptation to resources and data stream rates represent the first two stages. Merging the generated knowledge structures when running out of memory represents the last stage. AOG has been used in clustering, classification and frequency counting [15]. The function of the AOG algorithm is depicted in Fig. 3.

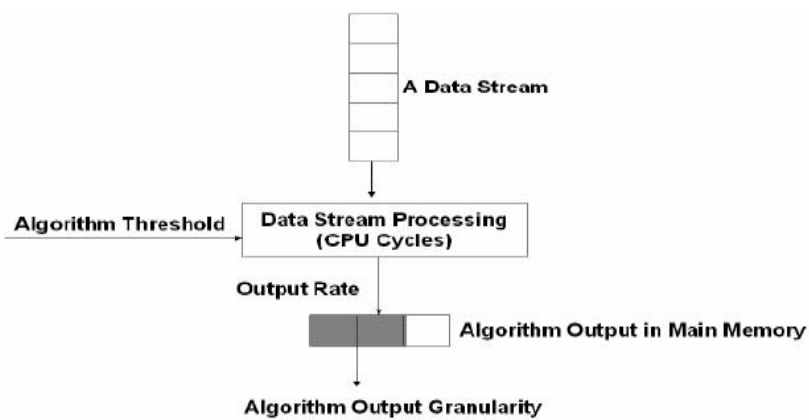

Fig. 3. The AOG algorithm [15]

## 3. Classification of data stream challenges

There are different challenges in data stream mining that cause many research issues in this field. Regarding to data stream requirements, developing stream mining algorithms is needed more studying than traditional mining methods. We can classify stream mining challenges in 5 categories; Irregular rate of arrival and variant data arrival rate over time, Quality of mining results, Bounded memory size and huge amount of data streams, Limited resources, e.g. ,memory space and computation power and To facilitate data analysis and take a quick decision for users. In the following each of them will be described.

One of the most important issues in data stream mining is optimization of memory space consumed by the mining algorithm. Memory management is a main challenge in stream processing because many real data streams have irregular arrival rate and variation of data arrival rate over time. In many applications like sensor networks, stream mining algorithms with high memory cost is not applicable. Therefore, it is necessary to develop summarizing techniques for collecting valuable information from data streams [19].

Data pre-processing is an important and time consuming phase in the knowledge discovery process and must be taken into consideration when mining data streams. Designing a light-weight preprocessing techniques that can guarantee quality of the mining results is crucial. The challenge here is to automate such a process and integrate it with the mining techniques.

By considering the size of memory and the huge amount of data stream that continuously arrive to the system, it is needed to have a compact data structure to store, update and retrieve the collected information. Without such a data structure, the efficiency of mining algorithm will largely decrease. Even if we store the information in disks, the additional I/O operations will increase the processing time. While it is impossible to rescan the entire input data, incremental maintaining of data structure is indispensable. Furthermore, novel indexing, storage and querying techniques are required to manage continuous and changing flow of data streams.

It is crucial to consider the limited resources such as memory space and computation power for reaching accurate estimates in data streams mining. If stream data mining algorithms consume the available resources without any consideration, the accuracy of their results would decrease dramatically. In several papers this issue is discussed and their solutions for resource-aware mining are proposed [9,14,16]. One of the proposed solutions is AOG which use a control parameter to control its output rate according to memory, time constraints and data stream rate [14, 16]. Also in [31] another algorithm is proposed that not only reduces the memory required for data storage but also retains good approximation given limited resources like memory space and computation power.

Visualization is a powerful way to facilitate data analysis. Absence of suitable tools for visualization of mining result makes many problems in data analysis and quick decision making by user. This challenge still is a research issue that one of the proposed approaches is intelligent monitoring [22].

We summarized these challenges and related research issues in Table 1.

Table 1. Classification of data stream mining challenges

| Research Issues | Challenges | Approaches |
|---|---|---|
| Memory management | Fluctuated and irregular data arrival rate and variant data arrival rate over time | Summarizing techniques |
| Data preprocessing | Quality of mining results and automation of preprocessing techniques | Light-weight preprocessing techniques |
| Compact data structure | Limited memory size and large volume of data streams | Incremental maintaining of data structure, novel indexing, storage and querying techniques |
| Resource aware | Limited resources like storage and computation capabilities | AOG and [31] |
| Visualization of results | Problems in data analysis and quick decision making by user | Still is a research issue (one of the proposed approaches is: intelligent monitoring) |

## 4. The proposed analytical framework

This research ends in an analytical framework which is shown in Table 2. This framework tries to show the efficiency of data mining applications in developing the novel data stream mining algorithms. These algorithms are classified base on the data mining tasks. We described the details of these algorithms based on preprocessing steps and the following steps. In addition, this framework can direct future works in this field.

Some of the most important results that have been reached during this research are:

(1) Mining data streams has raised a number of research challenges for the data mining community. Due to the resource and time constraints many summarization and approximation techniques have been adopted from the fields of statistics and computational theory.
(2) There are many open issues that need to be addressed. The development of systems that will fully address these issues is crucial for accelerating the science discovery in the fields of physics and astronomy, as well as in business and financial applications.

## 5. Conclusion

In this paper we reviewed and analyzed data mining applications for solving data stream mining challenges. At first we presented a comprehensive classification for data stream mining algorithms based on data mining applications. In this classification, we separate algorithms with preprocessing from those without preprocessing. In addition, we classify preprocessing techniques in a distinct classification. In the following, the layered architecture of the classification represents almost all of the challenges that are mentioned in various researches. Then we discussed the application of data mining techniques for addressing the challenges of data stream mining, and then we presented an analytical framework regarding these applications. Results are shown that it is necessary to adopt many summarization and approximation techniques from the fields of statistics and computational theory, besides crucial changes that are needed in common data mining techniques. In spite of the researches that have been done on data mining's application in data stream mining so far, there are still wide areas for further researches.

Table 2. Analytical Framework of Data Stream Mining Techniques

| | *Algorithm* | *Mining Task* | *Advantages* | *Disadvantages* |
|---|---|---|---|---|
| **Classification** | GEMM and FOCUS [17] | Decision tree and frequent item sets | - Concept drift detection<br>- Incremental mining models | - Time consuming and costly learning |
| | OLIN [24] | Uses info-fuzzy techniques for building a tree-like classification model | - Dynamic Update | - Low speed<br>- Storage memory problem<br>- Time consuming and costly learning |
| | VFDT and CVFDT [9] | Decision Trees | - High speed<br>- Need less memory space | - Non-adaption to concept drift<br>- Time consuming and costly learning |
| | LWClass [15] | Classification based on classes weights | - High speed<br>- Need less memory space | - Non-adaption to concept drift<br>- Time consuming and costly learning |
| | CDM [23] | Decision tree and Bayes network | - Suitable factor for distance measurement between events | - User defined information complexity |
| | On-demand stream classification [3] | Using micro-clusters ideas that each micro-cluster is associated with a specific class label which defines the class label of the points in it. | - Dynamic update<br>- High speed<br>- Need less memory space | - High cost and time need for labeling |
| | Ensemble-based Classification [32] | Using combination of different classifiers | - Single pass<br>- Dynamic update<br>- Concept drift adoption<br>- High accuracy | - Low speed<br>- Storage memory problem<br>- Time consuming and costly learning |
| | ANNCAD [25] | Incremental classification | - dynamic update | - Low speed<br>- Storage memory problem<br>- Time consuming and costly learning |
| | SCALLOP [13] | Scalable classification for numerical data streams. | - dynamic update | - Low speed<br>- Storage memory problem<br>- Time consuming and costly learning |
| **Clustering** | STREAM and LOCALSEARCH [28] | K-Medians | - Incremental learning | - Low clustering quality in high speed<br>- Low accuracy |
| | VFKM [11,12,21] | K-Means | - High speed<br>- Need less memory space | - Multi-pass |
| | CluStream [2] | The concepts of a pyramidal time frame in conjunction with a micro-clustering approach. | - Time and space efficiency<br>- concept drift detection<br>- High accuracy | - Offline clustering |
| | D-Stream [8] | Density-based clustering | - High quality and efficiency<br>- concept drift detection in real-time data stream | - High complexity |
| | AWSOM [29] | Prediction | - Efficient pattern detection<br>- Need less memory space<br>- dynamic update Single pass | - High complexity |
| | HPStream [9] | projection based clustering | - Efficient for high dimensional data stream<br>- Incremental update<br>- High scalability | - High complexity |
| **Frequency Counting and Time Series Analysis** | Approximate Frequent Counts [26] | Frequent item sets | - Incremental update<br>- Simplicity<br>- Need less memory space<br>- Single pass | - Approximate output with increasing error range possibility |
| | FPStream [18] | Frequent item sets | - Incremental and dynamic update<br>- Need less memory space | - High complexity |

**References**

[1] Aggarwal, C., Han, J., and Wang, J., (2004): *A Framework for Projected Clustering of High Dimensional Data Streams*. In Proceedings of International Conference on Very Large Data Bases. Toronto, Canada.
[2] Aggarwal, C., Han, J., Wang, J., and Yu, P.S., (2003): *A Framework for Clustering Evolving Data Streams*. In Proceedings of International Conference on Very Large Data Bases. Berlin, Germany.
[3] Aggarwal, C., Han, J., Wang, J., and Yu, P.S., (2004): *On Demand Classification of Data Streams*. In Proceedings of 2004 International Conference On Knowledge Discovery and Data Mining (KDD '04). Seattle, WA.
[4] Aggrawal, C.C. (2007). *Data Streams: Models and Algorithms*. Springer.
[5] Babcock, B., Babu, S., Datar, M., Motwani, R., and Widom, J., (2002): *Models and issues in data stream systems*. In Proceedings of the twenty-first ACM SIGMOD-SIGACT-SIGART symposium on Principles of database systems (PODS). Madison, Wisconsin, pp. 1-16.
[6] Babcock, B., Datar, M., and Motwani, R., (2003): *Load Shedding Techniques for Data Stream Systems*. In Proceedings of the 2003 Workshop on Management and Processing of Data Streams.
[7] Chaudhry, N.A., Show, K., and Abdelgurefi, M. (2005). *STREAM DATA MANAGEMENT, Advances in Database system*. Vol. 30. Springer.
[8] Chen, Y. and Tu, L., (2007): *Density-Based Clustering for Real-Time Stream Data*. In Proceedings of the 13th ACM SIGKDD international conference on Knowledge discovery and data mining. San Jose, California, USA, pp. 133-142.
[9] Chi, Y., Wang, H., and Yu, P.S., (2005): *Loadstar : Load Shedding in Data Stream Mining*. In Proceedings of the 31st VLDB Conference. Trondheim, Norway, pp. 1302-1305.
[10] Chu, F. (2005): *Mining Techniques for Data Streams and Sequences.* Doctor of Philosophy Thesis: University of California.
[11] Domingos, P. and Hulten, G., (2000): *Mining High-Speed Data Streams*. In Proceedings of the Association for Computing Machinery Sixth International Conference on Knowledge Discovery and Data Mining.
[12] Domingos, P. and Hulten, G., (2001): *A General Method for Scaling Up Machine Learning Algorithms and its Application to Clustering*. In Proceedings of the Eighteenth International Conference on Machine Learning: Morgan Kaufmann, pp. 106--113.
[13] Ferrer-Troyano, F.J., Aguilar-Ruiz, J.S., and Riquelme, J.C., (2004): *Discovering Decision Rules from Numerical Data Streams*. In Proceedings of the 2004 ACM symposium on Applied computing. Nicosia, Cyprus, pp. 649-653.
[14] Gaber, M.M., Krishnaswamy, S., and Zaslavsky, A., (2003): *Adaptive Mining Techniques for Data Streams Using Algorithm Output Granularity*. In Proceedings of the Australasian Data Mining Workshop.
[15] Gaber, M.M., Krishnaswamy, S., and Zaslavsky, A. (2006). *On-board Mining of Data Streams in Sensor Networks*, In *Advanced Methods of Knowledge Discovery from Complex Data*, S. Badhyopadhyay, et al., Editors., Springer. pp. 307-335.
[16] Gaber, M.M., Zaslavsky, A., and Krishnaswamy, S., (2004): *Resource- Aware Knowledge Discovery in Data Streams*. In Proceedings of First International Workshop on Knowledge Discovery in Data Streams. Pisa, Italy.
[17] Ganti, V., Gehrke, J., and Ramakrishnan, R. (2002): *Mining Data Streams under Block Evolution.* ACM SIGKDD Explorations Newsletter, **3** (2): pp. 1-10.
[18] Giannella, C., Han, J., Pei, J., Yan, X., and Yu, P.S. (2003). *Mining Frequent Patterns in Data Streams at Multiple Time Granularities*, In *Next Generation DataMining*, H. Kargupta, et al., Editors., AAAI/MIT,.
[19] Golab, L. and Özsu, M.T. (2003): *Issues in Data Stream Management.* ACM SIGMOD Record, **32** (2): pp. 5-14.
[20] Han, J. and Kamber, M. (2006): *Data Mining: Concepts and Techniques*. Second ed. The Morgan Kaufmann Series in Data Management Systems. Elsevier.
[21] Hulten, G., Spencer, L., and Domingos, P., (2001): *Mining Time-Changing Data Streams*. In Proceedings of the seventh ACM SIGKDD international conference on Knowledge discovery and data mining. San Francisco, California, pp. 97-106.
[22] Kargupta, H., Park, B., and Sarkar, K. (2002): *MobiMine: Monitoring the Stock Market from a PDA.* ACM SIGKDD Explorations Newsletter, **3** (2): pp. 37-46.
[23] Kwon, Y., Lee, W.Y., Balazinska, M., and Xu, G., (2008): *Clustering Events on Streams using Complex Context Information*. In Proceedings of the IEEE International Conference on Data Mining Workshops, pp. 238-247.
[24] Last, M., (2002): *Online Classification of Nonstationary Data Streams.* Intelligent Data Analysis, **6** (2): pp. 129-147.
[25] Law, Y. and Zaniolo, C., (2005): *An Adaptive Nearest Neighbor Classification Algorithm for Data Streams*. In Proceedings of the 9th European Conference on the Principals and Practice of Knowledge Discovery in Databases. Porto, Portugal: Springer Verlag.
[26] Manku, G.S. and Motwani, R., (2002): *Approximate frequency counts over data streams*. In Proceedings of the 28th International Conference on Very Large Data Bases. Hong Kong, China.
[27] Muthukrishnan, S., (2003): *Data streams: algorithms and applications*. In Proceedings of the fourteenth annual ACM-SIAM symposium on discrete algorithms.
[28] O'Callaghan, L., Mishra, N., Meyerson, A., Guha, S., and Motwani, R., (2002): *Streaming-data algorithms for high-quality clustering*. In Proceedings of IEEE International Conference on Data Engineering.
[29] Papadimitriou, S., Faloutsos, C., and Brockwell, A., (2003): *Adaptive, Hands-Off Stream Mining*. In Proceedings of the 29th International Conference on Very Large Data Bases VLDB.
[30] Tatbul, N., Cetintemel, U., Zdonik, S., Cherniack, M., and Stonebraker, M., (2003): *Load Shedding on Data Streams*. In Proceedings of the Workshop on Management and Processing of Data Streams. San Diego, CA, USA.
[31] Teng, W., Chen, M., and Yu, P.S., (2004): *Resource-Aware Mining with Variable Granularities in Data Streams*. In Proceedings of the 4th SIAM International Conference on Data Mining. Lake Buena Vista, USA, pp. 527-531.
[32] Wang, H., Fan, W., Yu, P., and Han, J., (2003): *Mining Concept-Drifting Data Streams using Ensemble Classifiers*. In Proceedings of the 9th ACM International Conference on Knowledge Discovery and Data Mining (SIGkDD). Washington DC, USA.